\documentclass[fleqn,twocolumn]{article}

\usepackage{amssymb}
\usepackage{amsmath}
\usepackage {graphics}

\title{\textbf{Polarization effects in the photon-induced process of\\
electron-positron pair creation  in a magnetic field, \\
studied in the ultra-quantum-mechanical approximation}}

\author{O. P. Novak \thanks{novak@ipfcentr.sumy.ua} \and R. I. Kholodov
\thanks{rkholodov@yahoo.com}}

\date{NATIONAL ACADEMY OF SCIENCES OF UKRAINE\\
Institute of applied physics\\
58, Petropavlivska St.,
40030 Sumy, Ukraine}

\begin{document}

\maketitle
\hrule height 1pt
\abstract{%
The photon-induced process of electron-positron pair creation in a strong
homogeneous magnetic field, provided that the polarization of particles is
arbitrary, has been considered. The polarization of a photon is described
in terms of the well-known Stokes parameters, and the relevant probabilities
of the process turn out to have simple analytical expressions, which allows
us to analyze the polarization and spin effects. A substantial influence of
the linear polarization of a photon on the spin orientations of electrons and
positrons has been demonstrated.
}
\vspace{0.5cm}
\hrule height 1pt


\section{Introduction}

Quantum-mechanical electrodynamic processes, which involve photons and
electrons in strong external electromagnetic fields, do not lose their urgency
for both experimental and theoretical studies, despite a good
number of available literature sources.

The process of creation of an electron-positron pair by a single photon in
a magnetic field has been considered for the first time by Klepikov
\cite{Klepikov} in the approximation of ultra-relativistic motion of particles.
The operator method for the consideration of this problem was applied by
Baier and Katkov in the quasiclassic ultra-relativistic case
\cite{Baier67, Baier68}. In recent years,
there has appeared the work of these authors \cite{BaierArXiv}, where the
operator method was used to study the process of electron-positron pair
creation by a photon,
with the particles being located at low-energy Landau levels. We
also mention work \cite{Semionova}, where the process of photon-induced
creation with the
participation of polarized particles has been considered for arbitrary Landau
levels and magnetic field values. Nevertheless, the complexity of general
formulas prohibited the authors from carrying out their detailed analysis.

A relatively little attention has been drawn to studying the polarization
effects in the given process in the case of nonrelativistic electrons in
weakly excited states, because the description of such a process requires
the application of the ultra-quantum- mechanical approximation, which, in its
turn, demands the presence of a rather strong magnetic field approaching the
critical Schwinger one $H_c = m^2c^3 / e\hbar = 4.41\times 10^{13}Gs$.

This work aims at calculating the probability of the photon-induced creation of
an electron-positron pair in a strong magnetic field and at elucidating the
roles of polarization and spin effects, as well as the correlations between
them. We found expressions for the probability of the process in a general
quantum-mechanical relativistic case and without putting any additional
confinements onto the values of such parameters as the momenta, energies,
field magnitude, and so on. The researches of similar expressions was carried
out by other authors \cite{Semionova}, but the
complexity of expressions prohibited them
from revealing the regularities associated with the polarization of the photon
and the spins of particles.

Owing to the specific choice of the approximation — low Landau levels and
magnetic fields with the magnitude below the critical one — we managed to
derive simple analytical expressions for the probability concerned
which explicitly depend on the Stokes parameters that
characterize the polarization of the photon.


\section{Probability of the process}
For the wave functions of an electron and a positron,
we used the following expressions (hereafter, we use the
system of units, where $\hbar = c =1$) \cite{Fomin2000}:

\begin{equation}
\label{eq1}
\begin{array}{l}
\displaystyle \Psi^-=\frac{1}{\sqrt S} e^{-i({E^-t-p^-_yy -p^-_zz})} A_{l^-}
\left[i\sqrt{{\tilde m^- - \mu^- m}} U_{l^-}(\zeta^-) + \right.\\
\displaystyle \left.\mu^- \sqrt{\tilde m^- + \mu^- m}
U_{l^- -1} (\zeta^-)\gamma^1 \right]u^-_{l^-} ,\\
\end{array}
\end{equation}
\begin{equation}
\label{eq1a}
\begin{array}{l}
\displaystyle\Psi^+= \frac{1}{\sqrt S} e^{i(E^+ t - p^+_y y - p^+_z z)}A_{l^+}
\left[ i\sqrt{\tilde m^+ + \mu^+ m} U_{l^+}(\zeta^+)- \right.\\
\displaystyle\left. \mu^+ \sqrt{\tilde m^+ - \mu^+ m}
U_{l^+ -1}(\zeta^+)\gamma^1 \right]u^+_{l^+}.
\end{array}
\end{equation}
Here, $S$ is the normalization area, $A_{l^\pm}$ are the
normalization constants, $p^\pm _y$ -- $y$-components of the
electron and positron momenta, $(\tilde m^\pm) ^2 = m^2 + 2l^\pm eH$,
\begin{equation}
\label{eq2}
\zeta^\pm = \sqrt{eH} \left(x - \frac{p^\pm_y}{eH} \right),
\end{equation}
$U_l(\zeta) = \frac{1}{\sqrt{\sqrt\pi 2^ll!}}
\exp(-\zeta^2/2)H_l(\zeta)$
is the Hermite function, $u^\pm_l $ are constant
bispinors, and $\mu^-$ and $\mu^+$ are the polarizations
of the electron and the positron, respectively.
For the wave function of the initial photon, we used
the standard expression \cite{RQT}
\begin{equation}
\label{eq3}
A = \frac{\sqrt{4\pi}} {\sqrt{2\omega V}}
\gamma^\mu e_\mu e^{-ikx},
\end{equation}
where $V$ is the normalization volume, $\gamma^\mu$ are the Dirac
matrices, and $e_\mu$ is the vector of photon polarization.

Let the vector of a uniform magnetic field be
directed along the axis $z$. Then, the wave functions
(\ref{eq1}), (\ref{eq1a}) are connected with the following gauge of the
electromagnetic potential:
\begin{equation}
\label{eq4}
A_0 = 0, \quad \vec{A} =(0,xH,0).
\end{equation}
In this case, the energy spectrum of the electron
(positron) looks like
\begin{equation}
\label{eq5}
(E^\pm _l)^2 =( p^\pm _z)^2 + (\tilde m^\pm )^2 =
(p^\pm _z)^2 + m^2 + 2l^\pm eH,
\end{equation}
where $l^\pm=0, 1, 2, \ldots$ is the principal quantum number
(the Landau level number) of the electron or positron.

Let us introduce the parameter $h = eH / m^2$ which is
the ratio between the magnetic field $H$ and the critical
field $H_c$. Thus,
\begin{equation}
\label{eq6}
\tilde m^\pm = \sqrt{1 + 2hl^\pm } .
\end{equation}

\begin{figure*}
\resizebox{\columnwidth}{!}{
\includegraphics{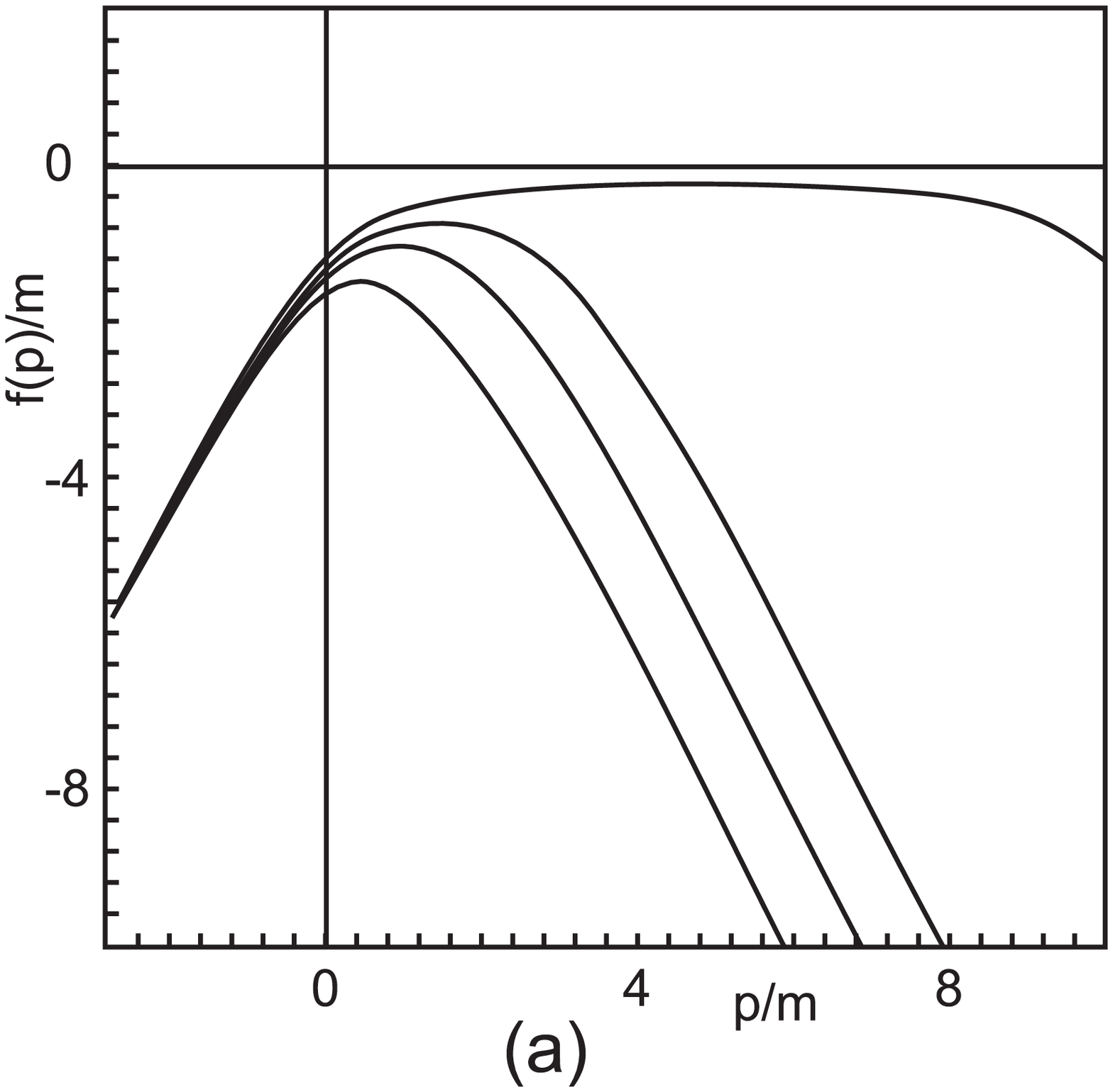}}%
\resizebox{\columnwidth}{!}{
\includegraphics{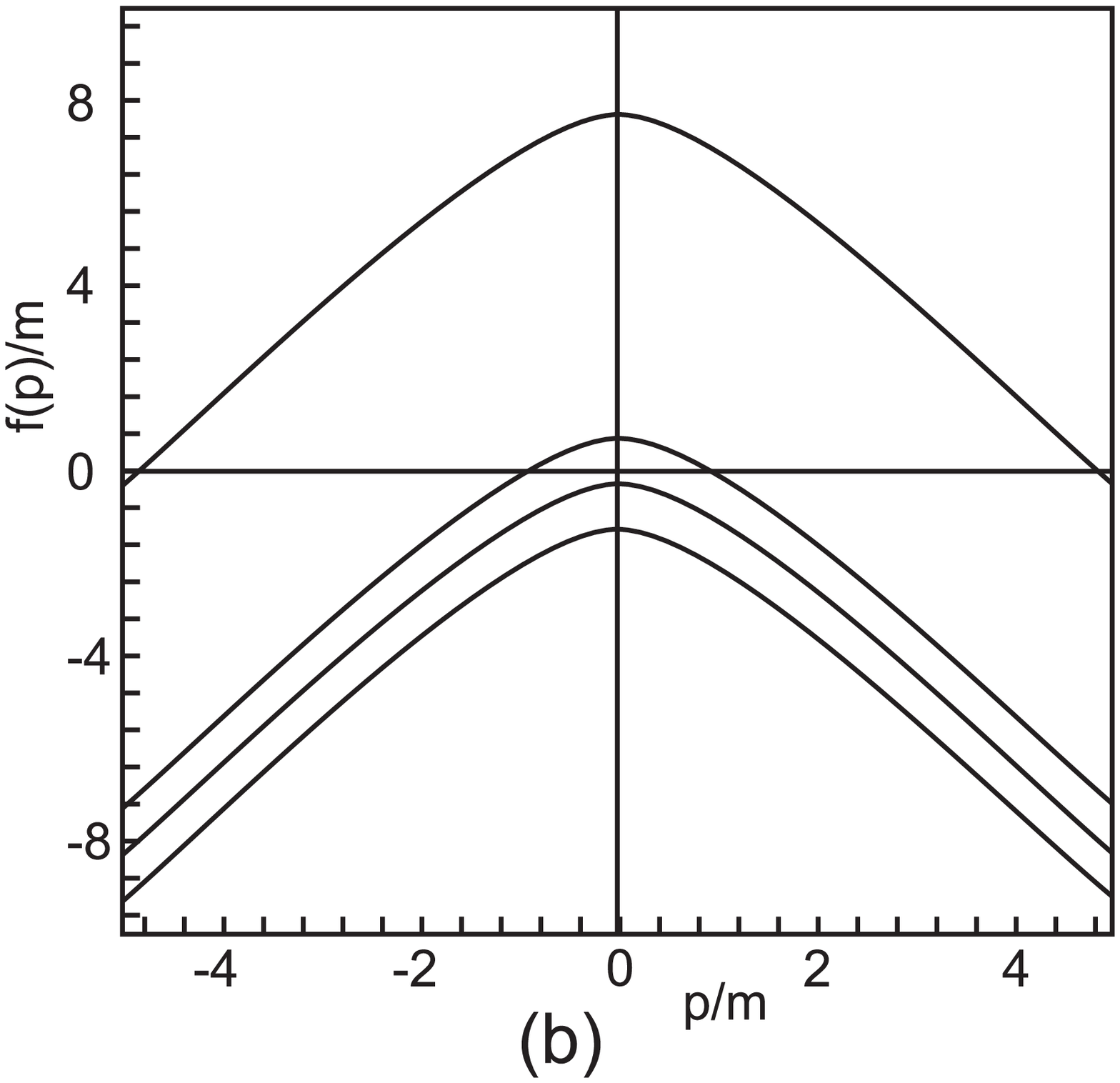}}
\caption{Plots of the function $f(p)$ at various frequencies of 1, 2, 3, and
10 (in terms of electron's mass units) and for the values $\theta=0$ (a) and
$\theta=\pi/2$ (b).}
\label{fig:1}
\end{figure*}

The following conservation laws are valid for the process
under consideration:
\begin{equation}
\label{eq7}
\left\{
\begin{array}{l}
 p^+_z + p^-_z = k_z=\omega\cos(\theta), \\
 E^- + E^+ = \omega , \\
 \end{array}
\right.
\end{equation}
where $\theta $ is the angle between the photon propagation
direction and the magnetic field. We are going to
determine the threshold values for the energy and the
pair momentum. Let us introduce the function
\begin{equation}
\label{eq8}
f(p) = \omega -
\sqrt{ ( \tilde m^- )^2 + p^2} -
\sqrt{ ( \tilde m^+ )^2 + (\omega u - p)^2 },
\end{equation}
where the notation $u=\cos(\theta)$ is used and -- for
convenience -- the subscripts for the $z$-components of the
electron momentum are omitted. The conservation laws
are fulfilled, provided the condition $f(p) = 0$ is true. The
threshold of the process is determined by the maximum
point of the given function. After differentiating, we find
\begin{equation}
\label{eq9}
f(p_m) = \frac{\omega (p_m - u E_m ) }{p_m}.
\end{equation}
The subscript $m$ denotes the threshold values. It is
obvious that the process is impossible, if the photon
propagates along the field, since, $u = 1$ in this case,
so that $E > p$ at any time (Fig. \ref{fig:1}, a). Figure \ref{fig:1}, b
demonstrates the plot of the function $f(p)$ for
$\theta= \pi / 2$.

By putting Eq. (\ref{eq9}) equal to zero, we determine the
required threshold values:
\begin{equation}
\label{eq10}
\omega _m = \frac{\tilde m^+ + \tilde m^-}{\sqrt{1 - u^2}},
\end{equation}
\begin{equation}
\label{eq11}
E^-_m = \frac{\tilde m^-}{\sqrt{1 - u^2}}  =
\frac{\tilde m^-}{\tilde m^- + \tilde m^+ }\omega _m \;,
\end{equation}
\begin{equation}
\label{eq12}
p^-_m = uE^-_m ,
\end{equation}
\begin{equation}
\label{eq13}
E_m^+ = \frac{\tilde m^+}{\sqrt{1 - u^2}} =
\frac{\tilde m^+}{\tilde m^- + \tilde m^+}\omega_m \;,
\end{equation}
\begin{equation}
\label{eq14}
p_m^+ = u E_m^+.
\end{equation}

To elucidate the physical meaning of the obtained expressions
(\ref{eq10}) -- (\ref{eq14}), we pass to a reference frame,
where $(p_m^-)'=0$. For this purpose, let us
write down the Lorentz transformation for the energy and the
momentum of the particle:
\begin{equation}
\label{eq15}
(p_m^-)' = \gamma (p_m^- - VE_m^-),
\end{equation}
\begin{equation}
\label{eq16} (E_m^ - )' = \gamma (E_m^ - - Vp_m^ - ),
\end{equation}
where $\gamma = 1/\sqrt{1 - V^2}$. Since
$(p_m^-)' = 0$, it follows from Eqs. (\ref{eq11}) and
(\ref{eq16}) that
\begin{equation}
\label{eq17}
V=u, \quad \gamma =\frac{\omega_m}{\tilde m^- + \tilde m^+},
\end{equation}
\begin{equation}
\label{eq18}
(E_m^-)'= \tilde m^-, \quad (E_m^+)' =\tilde m^+ ,
\end{equation}
\begin{equation}
\label{eq19}
(p_m^+)' = 0.
\end{equation}
Taking Eq. (\ref{eq7}) into account, we find that
\begin{equation}
\label{eq20}
\omega_m' = \tilde m^- + \tilde m^+,
\end{equation}
\begin{equation}
\label{eq21}
u' = 0.
\end{equation}

Thus, the threshold value for the photon frequency corresponds to the photon
frequency in the reference frame, where $\vec k \bot \vec H$, the
frequency is equal to the sum of the effective masses of electron and positron,
and the longitudinal momenta of particles are zero.

In the general case, the solution of Eq. (\ref{eq7}) gives
\begin{gather}
\label{eq22}
p^-_{1,2} = \frac{a u\pm b}{2\omega(1 - u^2)},\\
\label{eq23}
p_{1,2}^+ = \frac{a^+ u \mp b}{2\omega(1-u^2)},\\
\label{eq24}
E^-_{1,2} = \frac{a\pm bu}{2\omega(1 - u^2)},\\
\label{eq25}
E_{1,2}^+ = \frac{a^+ \mp bu}{2\omega(1-u^2)},\\
\end{gather}
where
\begin{gather}
\label{eq26}
a = \omega^2 (1-u^2) + (\tilde m^-)^2 - (\tilde m^+)^2,\\
\label{eq27}
a^+ = \omega^2 (1-u^2) - (\tilde m^-)^2 + (\tilde m^+)^2,\\
\label{eq28}
\begin{array}{l}
b^2 = a^2 - 4(\tilde m^-)^2\omega^2 (1-u^2) =\\
= (a^+)^2 - 4(\tilde m^+)^2 \omega^2 (1-u^2).
\end{array}
\end{gather}

Without loss of generality, we may assume that
\begin{equation}
\label{eq29}
u = \cos \theta = 0,
\end{equation}
because it is always possible to pass to the corresponding
reference frame, which moves along the magnetic field
and where equality (\ref{eq29}) is obeyed; the configuration
of the external field remains untouched at that.
For condition (\ref{eq29}), the following expressions for the
momenta can be obtained:
\begin{multline}
\label{eq30}
\displaystyle(p^-)^2= \frac{1}{4}\omega^2 - m^2 - (l^- + l^+)hm^2 +\\
\displaystyle +(l^- - l^+)^2 \frac{m^4h^2}{\omega^2} ,
\end{multline}
\begin{equation}
\label{eq31}
p^+ = -p^-.
\end{equation}

\begin{figure}[h]
\resizebox{\columnwidth}{!}{
\includegraphics{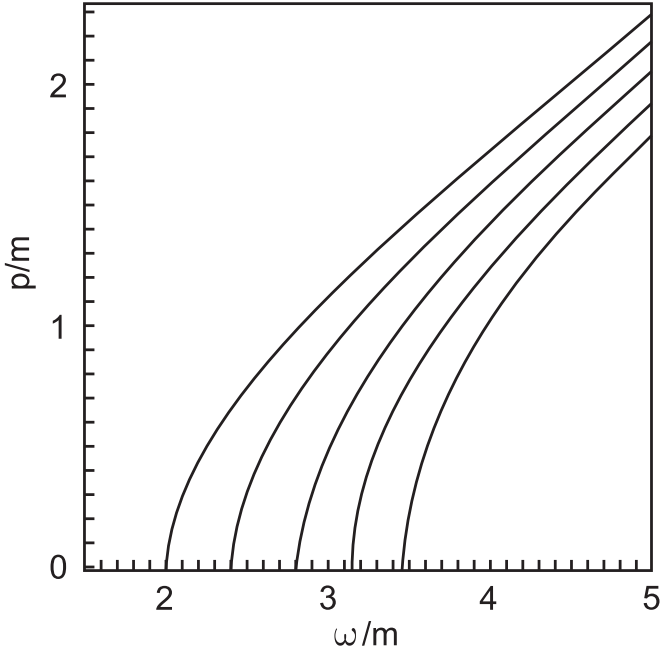}}
\caption{
Dependencies of the electron momentum on the frequency
of the initial photon (in terms of electron's mass units) for the
following Landau levels of the electron and the positron: (a)
$l^ - = 0$ and $l^ + = 0$, (b) $l^ - = 0$ and $l^ + = 5$, (c)
$l^ - = 5$ and $l^ + = 5$, (d) $l^ - = 5$ and $l^ + = 10$,
(e) $l^ - = 10$ and $l^ + = 10$.}
\label{fig:2}
\end{figure}

Formula (\ref{eq30}) is plotted in Fig. \ref{fig:2}.

The values of momenta of the created pair of particles substantially depend
on the frequency of the initial photon. In the case $\omega = \omega _m =
\tilde m^- + \tilde m^+$, the electron and the positron are created
accurately at the Landau levels, and $p_z^\pm = 0$. This gives rise, as is
known, to the emergence of divergences in the process amplitude.

The probability of the process, according to the well-
known rules of quantum electrodynamics \cite{RQT}, looks like
\begin{equation}
\label{eq32}
dW = \vert S_{fi} \vert^2 dN^- dN^+,
\end{equation}
where  $S_{fi}$ is the process amplitude;
\begin{equation}
\label{eq33}
dN^\pm = \frac{d^2p^\pm S}{(2\pi)^2},
\end{equation}
are the intervals of final states of the electron and the
positron, respectively; and $S$ is the normalization area.
According to general rules, the amplitude of the process
is determined -- in the first order of perturbation theory
-- by the formula
\begin{equation}
\label{eq34} S_{fi} = - ie\int {\bar {\Psi }^ - A^\mu \gamma _\mu
\Psi ^ + d^4x} .
\end{equation}
Considering the expressions for the wave functions of the
electron and the positron (see Eqs. (\ref{eq1}), (\ref{eq1a})), as well as
Eq. (\ref{eq3}), and carrying out the necessary integration procedure, we
obtain
\begin{equation}
\label{eq35} S_{fi} = \frac{\Phi \left( {2\pi } \right)^3\sqrt
{2\pi } }{4S\sqrt {\omega V} \sqrt {E^ - E^ + \tilde {m}^ - \tilde
{m}^ + } }A_{fi} \delta ^3\left( {p^ - + p^ + - k} \right),
\end{equation}
where
\begin{equation}
\label{eq36}
\begin{array}{l}
A_{fi} = \sqrt {\tilde {m}^ - - \mu ^ - m} \sqrt {\tilde {m}^ + +
\mu ^ + m}\times\\
\times R_+ \sin(\theta )\cos ( \alpha)J( {l^ + ,l^ - }) + \\
\vphantom{\sqrt{\sqrt{\tilde A}}}
+( - \mu ^+ )\sqrt {\tilde {m}^ - - \mu ^ - m} \sqrt{\tilde {m}^ + - \mu ^ + m}
R_ - T_ + J( {l^ + - 1,l^ - }) +\\
\vphantom{\sqrt{\sqrt{\tilde A}}}
+ \mu ^ - \sqrt {\tilde m^- + \mu^- m} \sqrt {\tilde m^+ + \mu^+m}
R_- T_- J(l^ + , l^ - - 1) +\\
\vphantom{\sqrt{\sqrt{\tilde A}}}
+( - \mu^- \mu^+)\sqrt {\tilde m^- + \mu ^ - m}
\sqrt {\tilde {m}^ + - \mu ^ + m} \times\\
\times R_ + \sin \left( \theta \right)\cos \left( \alpha
\right)J\left( {l^ + - 1,l^ - - 1} \right),
\end{array}
\end{equation}
and $\Phi$ is the phase factor. In formula (\ref{eq36}),
the following notations were
introduced:
\begin{equation}
\label{eq37}
R_ + = \frac{1}{R_m^ - R_m^ + }\left( {\left( {R_m^ +
} \right)^2p^ - + \left( {R_m^ - } \right)^2p^ + } \right),
\end{equation}
\begin{equation}
\label{eq38}
R_ - = \frac{1}{R_m^ + R_m^ - }\left( {\left( {R_m^ +
} \right)^2\left( {R_m^ - } \right)^2 - p^ + p^ - } \right),
\end{equation}
\begin{equation}
\label{eq39}
T_\pm = \cos \left( \theta \right)\cos \left( \alpha
\right)\pm ie^{i\beta }\sin \left( \alpha \right),
\end{equation}
\begin{equation}
\label{eq40}
R_m^\pm = \sqrt {E^\pm - \mu \: \tilde {m}^\pm } ,
\end{equation}
\begin{equation}
\label{eq41}
\tilde {m}^\pm = m\sqrt {1 + 2l^\pm h} .
\end{equation}

The special functions $J\left( {l^ + ,l^ - } \right)$ arise owing to the
integration of the amplitude over the coordinate $õ$, along
which the motion of a particle in the magnetic field
is quantized. The explicit expressions for the special
functions were taken from work \cite{Klepikov}:

\noindent (i) for $l^ - > l^ + $:
\begin{equation}
\label{eq42}
\begin{array}{l}
\displaystyle
J\left( {l^ + ,l^ - } \right) = e^{ - \frac{\eta
}{2}}\eta ^{\frac{\vert l^ - + l^ + \vert }{2}}\sqrt {\frac{l^ -
!}{l^ + !}} \frac{1}{\left( {l^ - - l^ + } \right)!}\times \\
\times F\left( { - l^ + ,l - l' + 1,\eta } \right);
\end{array}
\end{equation}

\noindent (ii) for $l^ - < l^ + $:
\begin{equation}
\label{eq43}
\begin{array}{l}
\displaystyle
J\left( {l^ + ,l^ - } \right) = e^{ - \frac{\eta
}{2}}\eta ^{\frac{\vert l^ - + l^ + \vert }{2}}\left( { - 1}
\right)^{l - l'}\times \\
\times \displaystyle
\sqrt {\frac{l^ + !}{l^ - !}} \frac{1}{\left( {l^
+ - l^ - } \right)!}
F\left( { - l^ - ,l' - l + 1,\eta } \right),
\end{array}
\end{equation}
where $F\left( {n,m,\eta } \right)$ is the degenerate hypergeometric
function. Its argument looks like
\begin{equation}
\label{eq44}
\eta = \frac{\omega ^2}{2m^2h}.
\end{equation}

We seek for the amplitude in the form
\begin{equation}
\label{eq45}
\begin{array}{l}
S_{fi} = \frac{1}{S\sqrt V }A\delta \left( {p_z^ - +
p_z^ + - k_z } \right)\times \\
\times \delta \left( {p_y^ - + p_y^ + - k_y } \right)\delta \left(
{E^ - + E^ + - \omega } \right),
\end{array}
\end{equation}
where
\begin{equation}
\label{eq46}
A = \frac{\Phi \left( {2\pi } \right)^3\sqrt {2\pi }
}{4\sqrt \omega \sqrt {E^ - E^ + \tilde {m}^ - \tilde {m}^ + }
}A_{fi} .
\end{equation}

According to formula (\ref{eq32}), the probability within the time unit is
\begin{equation}
\label{eq47}
\begin{array}{l}
dW = \frac{\vert A\vert ^2}{\left( {2\pi }
\right)^7}\frac{S}{V}\delta \left( {p_z^ - + p_z^ + - k_z }
\right)\times \\
\times \delta \left( {p_y^ - + p_y^ + - k_y } \right)\delta \left(
{E^ - + E^ + - \omega } \right)d^2p^ - d^2p^ + .
\end{array}
\end{equation}
Here, we used the following properties of the $\delta $-function:
\begin{equation}
\label{eq48}
\left( {\delta \left( {E^ - + E^ + - \omega }
\right)} \right)^2 = \frac{T}{2\pi }\delta \left( {E^ - + E^ + -
\omega } \right),
\end{equation}
\begin{equation}
\label{eq49} \left( {\delta \left( {p_i^ - + p_i^ + - k_i }
\right)} \right)^2 = \frac{L_i }{2\pi }\delta \left( {p_i^ - +
p_i^ + - k_i } \right),
\end{equation}
where the subscript $³$ designates the corresponding
coordinate ($x$ or $y$), and $L_i $ is the normalization distance.

After integrating over $d^2p^ + $ and $dp^ - _y $, and taking into
account that the integrand does not depend on $ð_{ó}$, we
find that
\begin{equation}
\label{eq50} dW = \frac{\vert A\vert ^2}{\left( {2\pi }
\right)^7}\frac{Sp^ - _y }{V}\delta \left( {E^ - + E^ + - \omega }
\right)dp^ - _z .
\end{equation}

As is seen, the factor $Sp_y / V$ appeared. To determine
its form, consider the argument of the wave function
(\ref{eq2}). Identifying the quantity  $p^ - _y / eH$ as a characteristic
distance of quantization $L_{x}$, which in terms of classical
physics has the meaning of the $x$-coordinate of the
Larmor orbit center, we obtain the relation
\begin{equation}
\label{eq51}
\frac{Sp^ - _y }{V} = \frac{p^ - _y }{L_x } = hm^2.
\end{equation}

Formula (\ref{eq50}) is the general expression for the probability
of the creation of an electron-positron pair with the
participation of polarized particles and without any
restriction put on the energies of those particles and the
magnitude of the magnetic field.

In what follows, we assume that $h<<1$. Moreover,
we use the so-called ultra-quantum-mechanical, or LLL
(Low Landau Levels) approximation:
\begin{equation}
\label{eq52}
l^\pm \sim 1
\end{equation}

We should emphasize that the ultra-quantum
mechanical case demands that a strong enough magnetic
field should be present. Really, even for an insignificant
transverse energy of an electron $E$ of about 10 keV, the
experimentally obtainable fields give $l$ of the order of $10^5$.
For conditions (\ref{eq52}) to be fulfilled, one needs such a magnetic
field which approaches the critical one by its order of magnitude.
Such fields are observed near pulsars.

Let the condition
\begin{equation}
\label{eq53}
hl^\pm < < 1.
\end{equation}
be fulfilled. In this approximation, we can determine the approximate values
of the momentum for the electron. Depending on the difference between the
frequency $\omega$ and the critical frequency $\omega _m $, expression
(\ref{eq30}) has the following forms in the ultra-quantum-mechanical
approximation:

(i) in the case where the additive to the critical frequency is of
the order of $h^2$,
i.e. $\omega = \omega _m + amh^2$, where $a$ is about unity,
the expression for
the momentum looks like
\begin{equation}
\label{eq54}
p_z^ - = \pm m\sqrt a h;
\end{equation}

(ii) if the condition $\omega = \omega _m + amh$ is satisfied,
the momentum will be equal to
\begin{equation}
\label{eq55}
p_z^ - = \pm m\sqrt {ah} ;
\end{equation}

(iii) at last, if the photon frequency considerably exceeds the critical one
for the selected Landau levels, i.e. $\omega = \omega _m + am$, we have
\begin{equation}
\label{eq56}
p_z^ - = \pm \frac{m}{2}\sqrt {a\left( {a + 4}
\right)} .
\end{equation}

Taking the aforesaid into consideration, we can expand Eq. (\ref{eq36})
in a power series in $h$. Let us write down the expressions
for the probability
of the pair creation process at the Landau level
$l^ - $,$ l^{ + }$ in the first
approximation. After integrating over $dp_z^ - $ we obtain (the superscripts
denote the polarization of the electron and the positron, respectively)
\begin{equation}
\label{eq57}
W^{ + + } = \frac{1}{4}\frac{\alpha m^4h^2}{\omega
E\vert p_z^ - \vert }Bl^ - \left( {1 - \xi _3 } \right),
\end{equation}
\begin{equation}
\label{eq58}
W^{ - - } = \frac{1}{4}\frac{\alpha m^4h^2}{\omega
E\vert p_z^ - \vert }Bl^ + \left( {1 - \xi _3 } \right),
\end{equation}
\begin{equation}
\label{eq59}
W^{ - + } = \frac{1}{2}\frac{\alpha m^4h}{\omega
E\vert p_z^ - \vert }B\left( {1 + \xi _3 } \right),
\end{equation}
where $\alpha $ is the fine-structure constant, $\xi _3 $ the Stokes parameter,
\begin{equation}
\label{eq60} \eta = \frac{\omega ^2}{2m^2h},
\end{equation}
\begin{equation}
\label{eq61} B = \frac{e^{ - \eta }\eta ^{l^ - + l^ + }}
{l^ - !l^+ !}.
\end{equation}

In the energetically unfavorable case $\mu ^ - = 1$ and $\mu ^ + =- 1$
the form of expression depends on the difference between the values of the
frequencies $\omega $ and $\omega _m $:

\noindent (i) in the case $\omega = \omega _m + amh^2$:
\begin{equation}
\label{eq62}
W^{ + - } = \frac{\alpha m^3h^5}{32\omega \vert p_z^
- \vert }Bl^ - l^ + \left( {\left( {1 + \xi _3 } \right) +
\frac{16\left( {p_z^ - } \right)^2}{m^2h^2}\left( {1 - \xi _3 }
\right)} \right);
\end{equation}

\noindent (ii) if $\omega = \omega _m + amh$:
\begin{equation}
\label{eq63}
W^{ + - } = \frac{\alpha mh^3}{2\omega }\vert p_z^ -
\vert Bl^ - l^ + \left( {1 - \xi _3 } \right);
\end{equation}

\noindent (iii) and for $\omega = \omega _m + am$:
\begin{equation}
\label{eq64}
W^{ + - } = \frac{\alpha m^4h^3}{4\omega E^3}\vert
p_z^ - \vert Bl^ - l^ + \!\left( {2\left( {1 - \xi _3 } \right) +
\frac{\left( {p_z^ - } \right)^2 \left( {1 + \xi _3 } \right)
}{2E^2}} \right).
\end{equation}

The total probability of the pair creation at arbitrary Landau
levels is given by the formula
\begin{equation}
\label{eq65} W_t^{\mu ^ + \mu ^ - } = \sum\limits_{l^ + ,l^ - }
{W^{\mu ^ + \mu ^ - }} .
\end{equation}

One can see that, depending on the polarization of created particles, the
value of the process probability can be characterized by different orders
of magnitude, which are determined by the degree of the small parameter $h$.
Highest is the probability of the process of pair creation in an energetically
profitable state, where $\mu ^ + = 1$ and $\mu ^ - = - 1$. Provided that
$h = 0.1$, it is by an order of magnitude larger that the others; therefore,
the total probability $W$ of the photon-induced creation of a pair with
arbitrary values of particle spin projections is governed just by this
summand:
\begin{equation}
\label{eq66}
W = \frac{1}{2}\frac{\alpha m^4h}{\omega E\vert p_z^
- \vert }B\left( {1 + \xi _3 } \right).
\end{equation}
Let us average this expression over the polarizations
of the initial photon. The photon polarization is described
by the quantum number $\lambda $ which accepts two values
$\lambda = \pm 1$. The change of the sign corresponds to the
sign change of the Stokes parameters. Thus, we obtain
\begin{equation}
\label{eq67}
 < W > = \frac{1}{2}\left( {W\left( { + \xi _3 } \right) + W\left( { - \xi
_3 } \right)} \right) = \frac{\alpha m^4h}{2\omega E\vert p_z^ -
\vert }B.
\end{equation}

\begin{figure}
\resizebox{\columnwidth}{!}{
\includegraphics{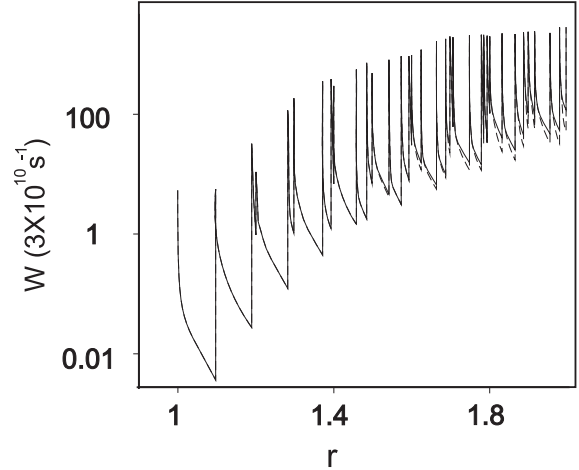}}
\caption{Dependence of probability (\ref{eq67}) summarized over
spins and averaged over photon polarization on the parameter
$r =\omega ^2 / 4m^2$. The dashed curve corresponds to the results
of work \cite{BaierArXiv}.}
\label{fig:3}
\end{figure}

One can see that the final formula is extremely simple. The plot of
dependence (\ref{eq67}) is depicted in Fig. \ref{fig:3}. The parameter
$r$ looks like
\begin{equation}
\label{eq68} r = \frac{\omega ^2}{4m^2}.
\end{equation}

It should be noted that our result obtained for small $l^ + ,l^ - $
is in accordance with
those obtained in works \cite{BaierArXiv, Semionova}. As the numbers
of Landau levels grow, a deviation of
our results from those of work \cite{BaierArXiv} is observed, which
is associated with the violation
of condition (\ref{eq52}) accepted by us.

Now, let us elucidate the origin of the resonance lines which are
observed in Fig. \ref{fig:3}. According to Eq. (\ref{eq50}), the general
formula includes the multiplier
$\delta \left({E^ - + E^ + - \omega } \right)dp_z $.
To integrate over $dp$, we must pass to the $\delta $-function depending
on momenta. As is known,
$\delta \left( {f\left( p \right)} \right) =
\sum {{\delta \left( {p - p_i } \right)}/{\left| {df / dp} \right|}} $,
where $p_i $ are the roots of the function $f\left( p \right)$. In our case,
$f\left( p \right) = E^ - + E^ + - \omega $. After the corresponding
differentiation, the function
$\delta \left( {E^ - + E^ + -\omega } \right)$ looks like
\begin{equation}
\label{eq69}
\delta \left( {E^ - + E^ + - \omega } \right) = \sum
{\frac{E^ - E^ + }{\left| {E^ - p^ + + E^ + p^ - } \right|}}
\delta \left( {p^ - - p_i^ - } \right).
\end{equation}

The denominator under the sum sign is zero, if $p^ - =p^+ = 0$,
i.e. if the pair becomes created accurately at the Landau level
and with zero longitudinal momenta.

From the physical point of view, the presence of singularities in
the probability of the pair creation is associated with the neglected
radiation emission of soft photons, which always accompanies
quantum- mechanical electrodynamic processes. This phenomenon
is similar to the so-called "infra-red catastrophe" that occurs in the course
of the bremsstrahlung process at the scattering by a Coulomb center
\cite{Fomin2007}.

\begin{figure}
\resizebox{\columnwidth}{!}{
\includegraphics{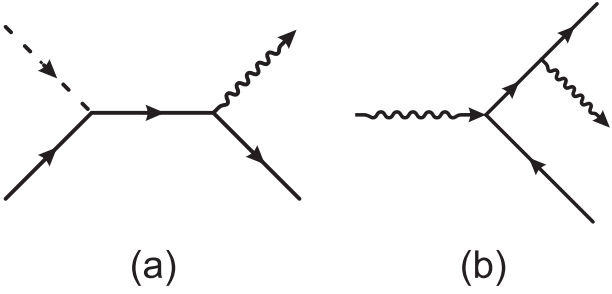}}
\caption{
Feynman diagrams of (a) bremsstrahlung and (b) photon-
induced pair creation with emission of a final photon.}
\label{fig:4}
\end{figure}

The process of electron scattering by a Coulomb center followed by
the emission of a photon is known to possess a similar divergence
(Fig. \ref{fig:4}, a). The cross-section of such a process
$d\sigma \sim {1}/{\omega '}$, where $\omega'$ is the final-photon frequency.
At $\omega ' \to 0$ we have $ d\sigma\to \infty $, which was named as the
"infra-red catastrophe". An analogous situation also takes place at the
photon-induced creation of the pair (Fig. \ref{fig:4}, b). Divergences arise,
because the perturbation theory becomes incorrect in the case of soft
photon emission.

\section{Polarization Effects}

The circular polarization $\xi _2 $ and the total linear polarization
$\sqrt {\xi _1^2 + \xi _3^2 } $ are known to be invariant with respect
to the choice of a coordinate system, whereas the parameters $\xi _1$
and $\xi _3$ to be not. It is convenient to determine the parameters
in that frame system, where their values are the same as those for
magnetic bremsstrahlung. Let the $x$- and $y$-axes lie in a plane which
is perpendicular to the wave vector $\vec {k}$; the $x$-axis being
oriented normally and the $y$-axis in parallel to the magnetic field
$\vec H$. In this case, the parameter characterizes the polarization
of the photon along the $x$-direction, and, provided the condition
$\xi_3 = - 1$, the photon becomes completely polarized perpendicularly
to the magnetic field. The parameter $\xi _1 $ determines the
polarization along those directions which are oriented at an
angle of 45$^{\circ}$ to the magnetic field.

It should be emphasized that the probability of the photon-induced pair
creation depends on the parameter $\xi _3 $ only, i.e. on the linear
polarization of the photon. This can be understood on the basis of
simple geometrical considerations. Owing to our choice of the
reference frame, where the condition $u =\cos \theta = 0$ is
satisfied, the vector of the electric field of the photon lies
in the same plane with the external field $\vec {H}$. Only the
component of the electric field that is normal to $\vec {H}$ and
lies in the plane of the classical orbit of the electron plays an
important role in the process. Therefore, only the linear polarization
can be included into the expressions for pair creation probability. The
probability cannot depend on $\xi _1 $, because, owing to the symmetry
of the problem, no direction that forms a 45$^\circ $-angle with
$\vec {H}$ is singled out.

Expressions (\ref{eq57}) -- (\ref{eq64}) readily demonstrate that there
is a clear correlation between the polarization of the photon and the
spin projections of the created pair. The processes with
($\mu ^ + = 1$,$\mu ^ - = 1$) and ($\mu ^
+ = - 1$, $\mu ^ - = - 1$) differ from the most probable one by the sign
of the initial photon polarization; the energetically unprofitable case
is characterized by a complicated dependence on the photon polarization,
which changes its shape with the variation of the ratio between the
transverse and forward energies of the created particles.

In probability (\ref{eq59}), we can select any value for the
polarization of the initial photon, because, actually, it
is determined by the experimental setup. In the case
$\xi _3 = - 1$, formula (\ref{eq59}) brings about the
zero probability of the process. Therefore, for a correct
comparison of probability values to be made in such a case,
it is necessary to determine the next correction with respect
to $h$ in Eq.~(\ref{eq59}). The corresponding calculation gave
rise to the expression
\begin{equation}
\label{eq70}
\begin{array}{l}
W^{ - + } = \frac{\alpha m^2B}{2\omega
}\frac{h}{g\varepsilon _0^5 }\left( {1 + \xi _3 } \right)\times \\
\times \left[ {\varepsilon _0^4 + \frac{1}{2}h\left( {3\varepsilon
_0^4 \left( {l^ - + l^ + } \right) - 2l^ - l^ + \varepsilon _0^2 }
\right)} \right],
\end{array}
\end{equation}
where the notations $g = {p_z^ - }/{m}$ and $\varepsilon _0^2 = 1 +
g^2$ were introduced. One can see that the character of the probability
dependence on the polarization is preserved, and, in the case $\xi _3 = - 1$,
the processes of creation of pairs with spin projections
($\mu ^ + = 1$, $\mu ^ - = 1$) and ($\mu ^ + = -
1$, $\mu ^ - = - 1$) really dominate.

Thus, by selecting the polarization of the initial
photon, one can influence the spin polarization of new
particles. As an example, we find the polarization degree
for electrons. By definition, it looks like
\begin{equation}
\label{eq71} P_{e^ - } = \frac{W^ + - W^ - }{W^ + + W^ - }.
\end{equation}
In our case, $W^ + = W^{ + + } + W^{ + - }$ and $W^ - = W^{ - + }
+ W^{ - - }$. Taking into account the fact that the probability
$W^{ + - }$  is by one order of magnitude lower than the other
ones, we can write
\begin{equation}
\label{eq72}
P_{e^ - } = \frac{W^{ + + } - W^{ - + } - W^{ - -
}}{W^{ + + } + W^{ - + } + W^{ - - }}.
\end{equation}
Depending on the $\xi _3 $-value, polarization (\ref{eq71}) is different.
If $\xi _3 \ne - 1$, the contribution $W^{ - + }$ exceeds all other
terms, so that, neglecting $W^{ + + }$ and $W^{ - - }$, we obtain
\begin{equation}
\label{eq73} P_{e^ - } \approx - 1.
\end{equation}
Hence, if the condition $\xi _3 \ne - 1$ is fulfilled, the beam of created
electrons will be almost completely polarized against the field direction.

Making use of Eqs. (\ref{eq57}), (\ref{eq58}) and (\ref{eq70}),
one can find a correction to $P_{e^ - } $. Let us substitute the
indicated expressions into Eq. (\ref{eq72}), expand the quantity
$P_{e^ - } $ in a power series in the small parameter $h$, and
confine ourselves to the first order of the expansion. After
simple calculations, we obtain
\begin{equation}
\label{eq74} P_{e^ - } = - 1 + hl^ - \frac{1 - \xi _3 }{1 + \xi _3}.
\end{equation}
But if $\xi _3 \to - 1$, then the quantity  $W^{ - + }$ in expression
(\ref{eq72}) can be neglected according to Eq. (\ref{eq70}), and we obtain
\begin{equation}
\label{eq75} P_ - = \frac{l^ - - l^ + }{l^ - + l^ + }.
\end{equation}

Therefore, if the condition $\xi _3\to - 1$ is obeyed, the polarization
degree of electron spins depends only on the numbers of Landau levels for
the electron and the positron. In the specific case $l^ + = l^ - = l$,
the degree of particle polarization is equal to zero. In such a case,
the quantity $P_{e^ - } $ is determined by terms that have higher order
with respect to the smallness parameter $h$; and the probability of the
process with $\mu ^ + = -1$ and $\mu ^ - = 1$ should be taken into account
in the calculation procedure. As is clear from
Eqs.~(\ref{eq62})--(\ref{eq64}), this probability has a simple form, namely
\begin{equation}
\label{eq76}
W^{ + - } = \frac{\alpha m^2}{\omega }\frac{e^{ -
\eta }\eta ^{l + l}}{\left( {l!}
\right)^2}l^2h^3\frac{g}{\varepsilon _0^3 }.
\end{equation}
under the conditions specified.

From Eqs. (\ref{eq77}) -- (\ref{eq79})  in the Appendix, we also obtain
that $W^{ + - } =
W^{ - + }$ and $W^{ + + } = W^{ - - }$; therefore, the numerator in Eq.
(\ref{eq71}) is equal to zero as well. Hence, if $\xi _3 \to - 1$
and $l^ + = l^ -$, the polarization degree of electrons vanishes
to within $h^3$.

Expressions (\ref{eq74}) and (\ref{eq75}) determine the polarization
degree of electrons in the case of their creation at the fixed Landau
levels $l^ +$ and $l^ - $. In order to find it in the case of the
photon-induced creation at arbitrary levels, the summation of
probabilities over  $l^ + $ and $l^ - $ has to be carried out
in relevant expressions.

\section{Conclusions}

In this work, the probability for the process of photon- induced
creation of an electron-positron pair in a strong magnetic field
has been found, taking into account the polarization of the
initial photon and the values of spin projections of created particles.
The total probability averaged over the photon polarization
(Eq. (\ref{eq67})) has a sawtooth dependence (Fig. \ref{fig:3})
and is in agreement with the results of other authors
\cite{BaierArXiv, Semionova}. The derived expressions
depend on the parameter of photon polarization through
the well-known Stokes parameters and have a simple
analytical form, which allowed us to carry out the
analysis of the polarization and spin effects.

A neat correlation was observed between the polarization of the photon
and the probability of the creation of particles with preset spin
projections. The corresponding probabilities depend on the linear
polarization $\xi_3$only; provided $\xi _3 \ne - 1$, the complete
polarization of particles' spins is observed. But $\xi _3 \to - 1$,
the degree of spin polarization depends only on the numbers of
Landau levels and is equal to zero in the specific case $l^ + = l^ - $.

\appendix

\section{Appendix}
For a more detailed analysis of polarization effects, it is
useful to know the expressions for the probability with a
higher accuracy with respect to the small parameter $h$.
Below, the results of calculations are given:
\begin{equation}
\label{eq77}
\begin{array}{l}
\displaystyle
W^{ - + } = \frac{\alpha m^2B}{2\omega}
\frac{h}{g\varepsilon _0^5 }  ( 1+\xi_3)
\left\{\vphantom{\frac{A}{A}} \varepsilon _0^4 + \right.\\
\displaystyle
+ \frac{1}{2}h \left[\vphantom{\tilde A}
3\varepsilon_0^4 (l^- + l^+) - 2l^- l^ + \varepsilon _0^2 \right]
-\frac{1}{4}h^2 \left[\vphantom{\tilde A}
\varepsilon_0^4 (l^2 + l'^2) + \right.  \\
\displaystyle  \left. \left.
+ll' \left( (l^- \!\! + l^+) (2\varepsilon_0^2 + \!\! 1)
- \! 2l^- l^+ \! - 8 - \! 12g^2 \! - 5g^4 \right)
\!\! \vphantom{\tilde A}\right]
\!\! \vphantom{\frac{A}{A}}\right\} +   \\
\displaystyle
+ \frac{\alpha m^2B}{4\omega }\frac{h^3g}{\varepsilon _0^3 }
\left( l^- + l^+ \right)^2,
\end{array}
\end{equation}
\begin{equation}
\label{eq78}
\begin{array}{l}
\displaystyle
W^{ + + } = \frac{\alpha m^2B}{4\omega
}\frac{h^2}{g\varepsilon _0^5 }\left( {1 - \xi _3 } \right)l^ -
\times \\
\displaystyle \times
\left\{ {\varepsilon _0^4 + \frac{1}{2}h\left[ {l^ - k_1 +
l^ + k_2 - 2l^ - l^ + \varepsilon _0^2 } \right]} \right\},
\end{array}
\end{equation}
\begin{equation}
\label{eq79}
\begin{array}{l}
\displaystyle
W^{ - - } = \frac{\alpha m^2B}{4\omega
}\frac{h^2}{g\varepsilon _0^5 }\left( {1 - \xi _3 } \right)l^ +
\times \\
\displaystyle\times
\left\{ {\varepsilon _0^4 + \frac{1}{2}h\left[ {l^ - k_2 +
l^ + k_1 - 2l^ - l^ + \varepsilon _0^2 } \right]} \right\},
\end{array}
\end{equation}
where
\begin{equation}
\label{eq80} k_1 = 3 - 2\varepsilon _0^3 + 4g^2 + g^4,
\end{equation}
\begin{equation}
\label{eq81} k_2 = 3 - 2\varepsilon _0^3 + 2g^2 - g^4.
\end{equation}

\section*{Acknowledgments}
The authors are grateful to P.~I.~Fomin for the formulation
of the problem
and his valuable remarks, as well as to S.~P.~Roshchupkin
for useful discussions.

\end{document}